\documentclass{doublecol-new}

\usepackage{natbib,stfloats}

\theoremstyle{TH}{

}

\theoremstyle{THrm}{

}

\theoremstyle{THhit}{

}

\makeatletter
\def\theequation{\arabic{equation}}
\makeatother

\usepackage{graphicx}

\usepackage{float}  
\floatstyle{ruled}
\newfloat{program}{thp}{lop}
\newfloat{programc}{thp}{lop}
 \floatname{program}{Program}
 \floatname{programc}{C Program}
\begin{document}%

\thispagestyle{plain}

\setcounter{page}{1}

\LRH{C.-Y. Wang et~al.}

\RRH{Arithmetic Operations Beyond 
Floating Point Number Precision}

\VOL{ }

\ISSUE{ }

\PUBYEAR{2011}

  \title{Arithmetic Operations Beyond Floating Point Number Precision} 

  \authorA{Chih-Yueh Wang* and Chen-Yang Yin}

  \affA{Department of Physics, Chung-Yuan Christian University \\
         200 Chung-Pei Road, Chung-Li, Taiwan 32023 \\
        Email: cw5b@msn.com; cywang@phys.cycu.edu.tw {*}Corresponding author} 



  \authorC{Hong-Yu Chen} %

  \affC{Institute of Astronomy, National Tsing Hua University, Hsinchu, Taiwan 30013}

  \authorD{Yung-Ko Chen} 

  \affD{National Synchrotron Radiation Research Center, Hsinchu, Taiwan 30076}

\begin{abstract}
In basic computational physics classes, students often raise the question of how to compute a number that exceeds the numerical limit of the machine. While technique of avoiding overflow/underflow has practical application in the electrical and electronics engineering industries, it is not commonly utilized in scientific computing, because scientific notation is adequate in most cases. We present an undergraduate project that deals with such calculations beyond a machine's numerical limit, known as arbitrary precision arithmetic. The assignment asks students to investigate the approach of calculating the exact value of a large number beyond the floating point number precision, using the basic scientific programming language Fortran. 
The basic concept is to utilize arrays to decompose the number and allocate finite memory.
Examples of the successive multiplication of even number and the multiplication and division of two overflowing floats are presented. 
The multiple precision scheme has been applied to hardware and firmware design for
digital signal processing (DSP) systems, 
and is gaining importance to scientific computing. 
Such basic arithmetic operations can be integrated to solve advanced mathematical problems to almost arbitrarily-high precision that is limited by the memory of the host machine.

\end{abstract}

\KEYWORD{computational physics education; floating point number; overflow/underflow; arbitrary precision; Fortran; GPU.}



\REF{to this paper should be made as follows: Wang,~C.-Y., Yin,~C.-Y., Chen,~H.-Y., Chen,~Y.-K. (2011) `Arithmetic Operations Beyond Floating Point Number Precision', {\it Int. J. Computational Science and Enginnering}, Vol.~, No.~, pp.~ ~ ~ ~ .}
%

\begin{bio}
Chih-Yueh Wang received a Ph.D. in Astronomy from the University of Virginia in 2001
and joined the faculty of CYCU in 2003. 
Her research interests are radiation hydrodynamic simulations of supernovae.  \vs{8}

\noindent 
Chen-Yang Yin graduated from CYCU in 2010 with a B.S. in Physics and is gallantly serving 1-year mandatory military service in the army.
\vs{8}

\noindent Hong-Yu Chen graduated from CYCU in 2009 with a B.S. degree in Physics and is working hard as an astronomy graduate student at National Tsing Hua University.
\vs{8}

\noindent Yung-Ko Chen graduated from CYCU in 2009 with a M.S. degree in electric engineering. He   
entered the National Synchrotron Radiation Research Center
in first place for 3-year substitutive military service. 
\end{bio}

 \maketitle\vfill\pagebreak
  \maketitle

\section{Introduction}

The concept of “number of significant figures” is important in various scientific and engineering disciplines. Fewer significant figures represents less precision, which can severely damage the result [1]. 
When calculations are made using computers, such that random and human errors do not apply, the highest and lowest values that can be represented and the precision of any number are limited by the finite number of digits that are used to store a floating point number, which is a binary version of a number in scientific notation, on the computer.

In the fixed point (integer) representation, a single precision integer is stored as 32 bits of 1s and 0s. Converted to a decimal value, such a number may be as low as $-2^{31}=-2,147,483,648$ or as high as $2^{31}-1=+2,147,483,647$. The precision of the significand (number of digits in the significand) is 32 bit, or ten digits in the decimal system. In the IEEE 754 standard for floating point arithmetic, a single-precision real number is also stored using 32 bits binaries , but one bit is used for the sign, eight bits store the exponent, and 23 bits store the significand of the number in scientific notation (called the fractional mantissa). The precision is 24-bit, and a decimal value between $\pm~1.4\times~10^{-45}$ and $\pm~3.4~\times~10^{+38}$ is allowed, with a significand precision of $log(2^{23})\approx~7$. Similarly, the IEEE 64-bit floating-point arithmetic standard provides 52 mantissa bits and 12 bits for the exponent, including the sign, giving approximately a 16 decimal digit accuracy and a decimal range of [$\pm~4.94\times~10^{-324},\pm~1.70\times~10^{308}$]. The length of the significand determines the precision with which numbers can be represented.

Since the number of binary bits that are used to express a number is finite, the set of numbers that can be stored by a computer, called machine numbers, does not cover all numbers in the physical world. Single-precision fixed point integers are precisely stored integers with a 32-bit resolution. Single-precision floats, however, are not exact representations with a  resolution of 24 bits; their value is truncated at the lower end, although numbers around $2^{127}$, much larger than the maximum of 32-bit integers, $2^{32}$, can be expressed. An overflow ($<-3.4\times~10^{+38}$ or $>3.4\times~10^{+38}$) or underflow  (between $-1.4\times~10^{-45}$ and $+1.4\times~10^{-45}$) error condition occurs when the number exceeds the maximum or falls below the minimum of machine numbers.

In some computations, even though the final number can be represented within machine limits, the intermediate results may overflow. For instance, the number of ways to select ten objects from a set of 200 is 200!/(190!10!)=$2.2\times~10^{16}$; the answer fits in the double precision range, but both 200! and 190! overflow. Although large factorials can be evaluated via Stirling's formula, which is used in formula calculators, such as the scientific calculator in Windows XP, in personal computers, the largest factorials that can be calculated as 32-bit and 64-bit integers are 12! and 20!, respectively. Additionally , the largest number that most electronic calculators using two-digit exponents can handle is 69! ($1.71\times~10^{98}<69!<10^{100}<70!$). Therefore, given available tools,
a simple combinatorial problem may become computationally formidable.

This study presents an undergraduate project on the methods for computing the exact value of a number beyond machine limits, known as arbitrary precision arithmetic or bignum arithmetic, using the high-level scientific programming language Fortran. While the approach has practical applications to the electric/electronic firmware design industry, it is not commonly adopted by physical scientists. The basic idea of this method  is to use arrays to store intermediate and final results, preventing numerical overflow or underflow. This investigation explores methods for multiplication and division.
Such basic arithmetic operations can be integrated to solve more complicated mathematical problems to a high precision that is limited by the memory of the host machine. 

\section{Arbitrary Precision Arithmetic using Arrays}

In a computer program, an array is an ensemble of numbers that represent data of the same class.
The size of the dataset (number of elements) can vary upon declaring the array. If a large or small number with many digits is divided into several segments, which are assigned to various elements of an array, then the number can be expressed without overflowing or underflowing the limits of the declared data type. For example, the 11-digit number `10123456789' can be stored in a single-precision integer array of 11 elements, with the ones digit `9' assigned to the first element of the array, the tens digit `8' to the second element, and so on. Each digit in the number is allocated to a single array element such that the whole number is represented by the array.

A single array element can also accommodate several digits, reducing computing time. If a 32-bit integer is to be represented, since the overflow threshold is $\sim 2 \times 10^{9}$, one array element can accommodate as many as ten digits. An array of 1000 elements can therefore represent a number of $\approx 10000$ digits, and if a 1000-digit number is to be stored, then an array with only 100 elements is needed. However, to prevent overflow in a subsequent computation, the maximum number of digits could  be optimized to be four or five , depending on the problem. The size of an array can be increased until the memory limit of the host computer is reached. Hence, the represented numbers are restricted only by the computer memory.

\section{Multiplication}

Consider the successive multiplication
$2\times 4 \times 6 \times 8 \times 10.... $ with a total of $N$ items,
such that the answer swiftly exceeds the upper limit of floating point numbers.
A general approach for avoiding overflow is to
take the logarithm of the expression and
then compute the exponent of the result.
However, logarithms and exponentiation introduce rounding errors,
so this method can not yield the exact answer.

The unknown is here split into the multiplication of two terms:
a power of two $2^N$ and a factorial $N!$.
In the program ``successive\_multipl'',
$M^N$, where $M=2$,
is first calculated. The subroutine
``multip\_parameter\_reset''
assigns initial values to the array elements $A(i)$; all elements are zero except that $A(1)=1$.
In the subroutine ``multipl'',
the initial multiplicand $A(1)=1$ is multiplied by the multiplier $M$
($M=2$ for computing $2^N$),
 and the result is stored back to $A(1)$.
Parameter $C$ denotes whether extra digits are produced.
If the result has more than 1 digit ($C/=0$), then the extra digit is carried to (added to) the next product
and array element $A(2)$.
Take $58 \times 3$ for example,
$8\times 3=24$ is first calculated, and the ones digit `4' is assigned to
$A(1)$.
The tens digit `2' is carried to the tens place,
such that $A(2)$ is assigned with `7', which is the ones digit of the 2nd (i=2) operation, $5\times 3 +2=17$.
In the 3rd (i=3) operation, the tens digit `1' of the 2nd operation is passed over to the hundreds place,
such that $A(3)$ stores the ones digit of $0 \times 3 + 1 \rightarrow = 1$.
The same procedure is repeated for each digit (the $i$ loop in ``multipl'') and each multiplication (the $i$ loop in the main program).
In program ``successive\_multipl'', 
each element of the array $X$ stores one digit of the final result of $2^N$. 
 The operation is similar to written multiplication based on decomposing the multiplicand.
The $k$ loop serves to carry a multi-digit number. It can be omitted in the calculation of $2^N$, because only a single-digit number has to be carried forward.

The method to compute the factorial of $N$ is the same, except that
the multiplier is replaced by $M=M+1$.
The array $Y$ stores the final result of $N!$. 
In program ``successive\_multipl'' (program 1 and 2), the power term and the factorial are multiplied together to attain the answer in the subroutine ``multiply\_two\_Ans''.
Each digit of the multiplicand $X(i)$ is multiplied by the that of the multiplier $Y(i)$.
The intermediate products
$P(i+j-1)=X(i)*Y(j)$
are added up and accumulated in columns.
Then
in the subroutine ``multipl'',
extra digits in a product are carried to next higher places and the numbers are rearranged,
which computes the product of $P$ and $1$ ($M=1$).
Figure 1(a) illustrates the decomposed method used in ``multiply\_two\_Ans''.
The total number of digits  
of the multiplied result
is either $Nx+Ny-1$ or $Nx+Ny$, where $Nx$ and $Ny$ are the numbers of digits in $X$ and $Y$.
Notably, the answer to $2 \times 4 \times 6 \times 8.....$ can also be computed directly via the subroutine ``multipl'',
 with $M=M+2$ being the multiplier.


The algorithm program 2  
and Fig. 1(b) present the second algorithm for multiplication. 
This method differs from the previous one in that a double loop is used,
 such that the product $X(i-1)Y(j)$  is carried to the next one $X(i)Y(j)$ immediately after each element is attained. 
Intermediate products, i.e., remainders,
are attained in a row.
 The same operations is applied for various $j$ to yield another row of remainders,
which values are added to previous ones in columns once they are obtained.
Iterations of the above steps yield the final product and remainders (denoted by $P(i+j)$ or $P(k)$). 
 The multiplication of overflowing or underflowing fixed numbers can be extended straightforwardly to that of floating-point numbers provided the exponents of the fixed numbers are specified.
Program ``multi2'' (program 4) shows the multiplication of two floats. 

Figure 1(c) and program ``multi3\_reverseorder'' suggest a similar method to method 2. 
Each digit of the number is reversely ordered
such that the highest place of a number is stored in the first array element.
In each $i$ loop the $X(i)Y(j)$ products are similarly carried leftward in a row to compute the intermediate product (remainder).  
 Contrary to method 2, 
remainders in various rows (varying $i$) are summed in columns
only when all of the remainders have been attained.
In the computer program the summation is accomplished via the use of a separate double do-loop.

The memory required for such arithmetic operations is well below the typical size of memory adopted in home computers. Notably, a single-precision integer occupies four bytes of memory. If each single-precision array element stores one digit, then such an array of 200,000 elements can represent 0.2 million digits,
whose text output is about $\sim 600$ pages long, 
at the expense of only 1MB of memory. 
Both the long program, invoking ``multiply\_two\_Ans'' , and the direct method, using only ``multipl'', take $\sim 150$ s to compute the 228288 digit-long value of $50000! \times 2^{50000}$.

    \begin{program}
 \begin{footnotesize}
\begin{verbatim}
module global
   implicit none
   integer, parameter :: s=230000   ! array size 
   integer :: A(s),X(s),Y(s),P(s),M,C,Ntot,N=50000 
end module

PROGRAM SUCCESSIVE_MULTIPL
   use global
   implicit none
   integer :: i, Nx, Ny 
   real :: ti, tf
   CALL CPU_TIME(ti)

! -----calculate 2^N------ 
   CALL multipl_parameter_reset  
   do i = 1, N
      M = 2
      CALL multipl  ! 2^N
   end do
   X  = A
   Nx = Ntot
   write(*,"(6H2^N = ,230000I1)") (X(i), i=s,1,-1)
   print*,"Number of digits of 2^N =", Nx

! -----calculate N!-----
   CALL multipl_parameter_reset  
   do i = 1, N
      M = M + 1
      CALL multipl   ! N!
   end do
   Y  = A
   Ny = Ntot
   write(*,"(6HN!  = ,230000I1)") (Y(i), i=s,1,-1)
   print*,"Number of digits of N! =", Ny

! -----calculate 2^N * N!------
   CALL multiply_two_Ans(Nx, Ny)  
   P  = A
   write(*,"(13H(2^N)*(N!) = ,230000I1)") (P(i), i=s,1,-1)
   print*,"Number of total digits =", Ntot
   CALL CPU_TIME(tf)
   print *, 'computing time (sec) = ', tf-ti
END PROGRAM successive_multipl

SUBROUTINE multipl_parameter_reset
   use global
   implicit none
   M   = 0
   C   = 0
   Ntot = 1
   A    = 0 
   A(1) = 1
end Subroutine

SUBROUTINE multipl
use global
   implicit none
   integer :: i, k 
   do i = 1, Ntot
      A(i) = A(i)*M + C
      C = A(i)/10
      A(i) = mod(A(i),10)
   end do
   if (C /= 0) then
       Ntot =  log10(dble(C) )  + Ntot + 1
       do k = i,Ntot
          i = Ntot
          A(k) = A(k)+ mod(C,10)
          C = C/10
       enddo
   end if
END SUBROUTINE
\end{verbatim}
\caption{Successive Multiplication of Even Numbers}
 \end{footnotesize}
\end{program}

    \begin{program}
\begin{footnotesize}
\begin{verbatim}
SUBROUTINE multiply_two_Ans(Nx,Ny)
use global
   implicit none
   integer :: i, j, Nx, Ny
   do j = 1, Ny
      do i = 1, Nx
         P(i+j-1) = P(i+j-1) + X(i)*Y(j)
      end do
   end do 
   CALL multipl_parameter_reset
   A = P
   M = 1
   Ntot = Nx + Ny - 2
   CALL multipl
END SUBROUTINE
\end{verbatim}
\caption{Successive Multiplication of Even Numbers - Continued }
\end{footnotesize}
\end{program}

  \begin{program}
 \begin{footnotesize}
\begin{verbatim}

! initialization
  K = 4;        // number of digits for an array element
  L = 10 ** K;  // number limit for an array element
  for i = 0 to Nx+Ny
     P(i) = 0;
  end;

! multiplication
  for j = 1 to Ny
     for i = 1 to Nx 
       T = X(i) * Y(j) + P(i+j-1);
       C = T / L;  // number to be carried to next digit
       P(i+j-1) = T % L = T - C * L;  
                   // accumulated i+j-1 element of product
       P(i+j) = P(i+j) + C;
     end;
  end;
\end{verbatim}
\caption{Algorithm:  Multiplication $X \cdot Y$ - Method 2}
 \end{footnotesize}
\end{program}

  \begin{figure}
\centering
(a)
   {
              \includegraphics[width=2.4in]{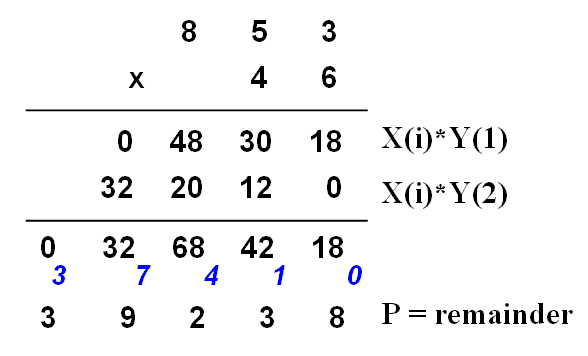} 
\vskip 0.3in
   }
(b)
    {
        \includegraphics[width=2.4in]{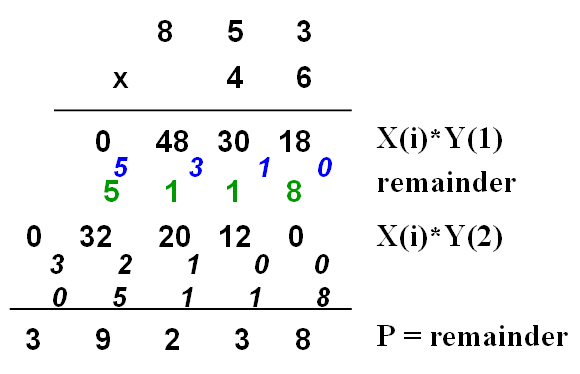}   
\vskip 0.3in
    }
(c)
    {
       \includegraphics[width=2.4in]{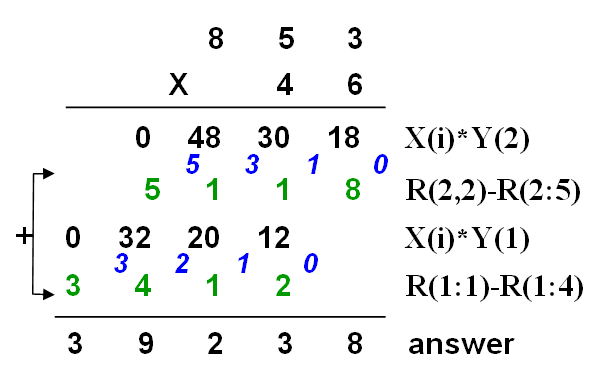} 
\vskip 0.3in
    }
   \vskip 0.3in
   \caption{Three decomposed methods for multiplication.}
\end{figure}

   \begin{program}
\begin{footnotesize}
 \begin{verbatim}

PROGRAM multi2 

  implicit none
  integer :: X(5)=0, Y(4)=0, P(9)=0
  integer :: T, C, i, j, k, L, Ex, Ey, Etot 
 
  X(5:1:-1) = (/ 8,5,3,4,6 /)    
  Ex = -3                        

  Y(4:1:-1) = (/ 7,1,2,9 /)      
  Ey = +2 

  L = 10

  do j = 1, 4 
     do i = 1, 5 
        T = X(i)*Y(j) + P(i+j-1)
        C = T / L
        P(i+j-1) = mod(T,L) 
        P(i+j) = P(i+j) + C
     enddo
  enddo

  Etot = Ex + Ey

  print"(1I1, 1H. 8I1, 3H E+, I3)", (P(k), k=9,1,-1), Etot+9-1

END PROGRAM

$ a.out
6.08431634 E+  7
\end{verbatim}
\caption{Multiplication of Two Floats - Method 2}
 \end{footnotesize}
\end{program}

  \begin{program}
 \begin{footnotesize}
\begin{verbatim}

PROGRAM multi3_reverseorder

  implicit none
  integer :: i, j, k, T, L=10, m
  integer :: C, C2  ! numbers to be carried when summing 
                    ! remainders in rows/columns 
  integer, parameter :: Nx= 5, Ny= 4, Ntot=Nx+Ny ! array size
  integer, parameter :: Ex=-3, Ey=+2, Etot=Ex+Ey ! exponents
  integer :: X(Nx)=0, Y(Ny)=0, R(Ny,Ntot)=0, P(Ntot)=0

  X(1:Nx) = (/8, 5, 3, 4, 6/)    
  Y(1:Ny) = (/7, 1, 2, 9/)       

  m = Ny

AA: do j = Ny, 1, -1   
       C=0
  BB:  do i = Nx, 1, -1
          T = X(i)*Y(j) + C
          C = T / 10
          R(j, i+m) = mod(T,L)
          if (i==1) then
              R(j, m) = C
          endif
       enddo BB
       m = m - 1
    enddo AA

    C2=0

CC: do k = Ntot, 1, -1
       P(k) = P(k) + C2
  EE:  do j = Ny, 1, -1
          P(k) = R(j,k) + P(k)
       enddo EE
       C2 = P(k) / 10
       P(k) = mod(P(k),L)
    enddo CC 

  write(*, "(9I1,1X,4H*10^,1X,I3)") P, Etot 
  print"(1I1,1H. 8I1,3H E+,I3)",(P(k),k=1,Ntot),Etot+Ntot-1

END PROGRAM
\end{verbatim}
\caption{Multiplication of Two Floats - Method 3}
 \end{footnotesize}
\end{program}

\section{Division}

\subsection{Simple Division}

Division may yield an infinite decimal without recurrence. The precision of division may therefore depend on the declared array size of the quotient. Double precision floats only allow a maximum of 15-16 digits in the significand. The algorithm program 6 
illustrates how to acquire a higher precision, say 30 digits, to the right of the decimal point, for a simple division of two numbers within the machine limits, $Q=X/Y$.

The integer quotient $Q0$ is obtained first. The remainder is stored in the variable $R$. The remainder is multiplied by 10, which result is taken as the new dividend and is divided by the divisor. The  first decimal place of the quotient is then obtained, and assigned to the first element of array $Q(i)$. The same operations are performed for each digit 
until the required digits have been computed and stored. The idea is the same as that of division by hand in which the dividend is decomposed. 

  \begin{program}
\begin{footnotesize}
\begin{verbatim}

  Q0 = X / Y       // integer quotient
  R  = X % Y       // remainder

  for i = 1, array_size 
      Q(i) = R * 10 / Y ;  // decimal quotient 
      R    = R % Y ;
  end ; 
\end{verbatim}
\caption{ Algorithm: Simple Division}
  \end{footnotesize}
\end{program}

\subsection{Successive Division}

Division of an arithmetic progression of numbers, $1/2/4/6/8/10...$ with a total of $N$ items is considered in program ``sd03''. All of the temporary dividend and quotient $T(i)$ are initially `0' , except that $T(1)=X=1$. When the nth decimal place of the quotient is attained, it is stored in the (n+1)th element of $T$. The final decimal quotient $Q(1:s)$
is extracted from $T(2:s+1)$. Since the value that is assigned to each array element is restricted by the machine number,  in any step, the new dividend $10\times R$ cannot be larger than the limit of a 32-bit fixed point number, $2,147,483,647$, and the divisor cannot exceed $2147483647/10$, as in that case overflow would occur.
   \begin{program}
 \begin{footnotesize}
 \begin{verbatim}

PROGRAM sd03  

  implicit none
  integer, parameter :: s = 1000 ! array size
  integer :: X = 1, Y = 1, R
  integer :: Q(s), T(s+1)
  integer :: i, j, N=99 

  T = 0
  T(1) = X
  
  do j = 1, N

    Y = Y+1   ! arithmetic progression
!   Y = Y*2   ! geometric progression

    do i = 1, s
       R    = T(i)*10 + T(i+1)
       Q(i) = R/Y
       R    = mod(R,Y)
       T(i+1) = R
    end do 

    T(1) = 0
    T(2:s+1)=Q(1:s)

  end do

  write(*,"('0.',1000I1)") Q

END PROGRAM sd03
\end{verbatim}
\caption{ Successive Divisions of Even Numbers}
 \end{footnotesize}
\end{program}

\subsection{Division of Two Overflowed Floats}

In the division of two long overflowing numbers, each number is partitioned into one array element, but in reverse order, such that the highest place of each number is stored in the first element. 
The basic idea for division involves subtraction that is carried out repeatedly to compute the quotient and remainder.

Consider 388756/129. From the left of the dividend, take as many digits as necessary to form a number [3,8,8] that contains the divisor [1,2,9] at least once but fewer than ten times. Subtract the divisor [1,2,9] from the partial dividend [3,8,8] to obtain the first partial remainder [2,5,9]. Repeat the subtraction (388-129=259, 259-129=130, 130-129=001, 001-129=-128) until a remainder close to zero,  [0,0,1], is reached (Fig. 2(a)). If the remainder becomes negative, then add the divisor back to restore the subtraction. These repeated subtractions yield a result that is identical to that of dividing the partial dividend by the divisor once. The number of subtractions performed on the partial dividend [3,8,8] is the partial quotient at the highest place. Accordingly, the number 3 is stored as the first element of the quotient: Q(1)=3.

To compute the next digit of the quotient, take one or more digits starting from the remainder’s places in the dividend
to form a new partial dividend that is larger than the divisor. Subtract the new partial dividend repeatedly until a partial quotient is yielded. Here, two more digits are included to form the new partial dividend [1,7,5],
so the next partial quotient Q(2)=0. [1,7,5] is subtracted by [1,2,9] once and Q(3)=1 is yielded. The remainder is 46.

In the third cycle, [1,2,9] is subtracted from [4,6,6] four times; therefore, the partial quotient is Q(4)=4.
In the fourth cycle, since the last digit of the dividend has been reached, `0' is appended to the remainder 79 to form a partial dividend [7,9,0]. The operations are repeated  until all of the digits of the quotient have been computed.

Program ``division01''          
performs the division of two floating-point numbers partitioned into arrays.
The values expressed by the arrays can be viewed either as integers of floats bearing
the same number of digits to the left of the decimal point,
 while $Ex$ and $Ey$ denote the exponents.
The zeroth elements are initialized to zero. Parameters `Li' and `Lf' refer to the initial and final places of the dividend that is used to form a partial dividend. Initially, Li=1 and Lf=Ny=3.
In the second cycle, since the partial dividend begins from two digits to the right, `Li=3' and `Lf=6'. The operations terminate when 
the quotient is expressed to the $s-1$ decimal places,
or $Lf-Ny+1==s$, where $s$ is the user-supplied array size.

The first loop of the program illustrates the subtraction of a set of array elements from another set of such elements. If the value of the dividend element is smaller than that of the divisor element, such as in  [3,8,8]-[1,2,9], where $X(i)=8<Y(j)=9$,
then 10 is borrowed from the next higher digit $X(i-1)$ and added to $X(i)$, and 1 is subtracted from $X(i-1)$, such that the subtraction becomes [3,7,`18']-[1,2,`9']=[2,5,`9'].
If the higher place becomes negative ($X(Li-1)<0$), as in  [0,0,1]-[1,2,9], the new remainder is negative and the subtraction method fails. The subtraction must be restored by resetting the dividend $X$ to the older remainder $R$ attained in the preceding cycle.

    \begin{program}
 \begin{footnotesize}
 \begin{verbatim}

PROGRAM division01

  implicit none
  integer :: i, j
  integer, parameter :: ss=200, s=30 
  integer, dimension(0:ss) :: R=0, X=0, Y=0
  integer, dimension(1:s)  :: Q=0
  integer, parameter :: Nx=6, Ny=3, Ex=2, Ey=-4, Etot=Ex-Ey
  integer :: Li=1, Lf=Ny  

  X(0:Nx) = (/0,3,8,8,7,5,6/) ! Ex=+2, X = 3.88758E+2
  Y(0:Ny) = (/0,1,2,9/)       ! Ey=-4, Y = 1.29E-4

99  R = X

!   Array subtraction: subtract divisor from partial dividend.
AA: do i = Lf, Li, -1   
       j = i - Lf + Ny    
       if (X(i) < Y(j)) then  
           X(i) = X(i) + 10         ! borrow 10 from left  
           X(i) = X(i) - Y(j)       ! subtract divisor 
           X(i-1) = X(i-1)-1        ! subtract 1 from left
       else if (X(i) >= Y(j)) then
           X(i) = X(i) - Y(j)       ! subtract directly
       end if  
    end do AA

!   Add numbers of subtraction to partial quotient.
    Q(Lf-Ny+1)=Q(Lf-Ny+1)+1

!   Restore subtraction when remainder becomes negative.
    if( X(Li-1) < 0 ) then  
        Q(Lf-Ny+1) = Q(Lf-Ny+1) - 1
        Li = Li + 1  
        Lf = Lf + 1  
        X = R   
    end if
 
!   Repeat operations.
    if( Lf-Ny+1 /= s ) goto 99 

  print*, " quoitent:"
  write(*, "(1I1, 1H., 29I1, 7H * 10^ , 1I5)") &   
        Q(1), (Q(i), i =2, s), Etot

END PROGRAM

$ a.out
  quotient:
3.01361240310077519379844961240 * 10^     6
\end{verbatim}
\caption{Division of Two Floats - Method 1}
 \end{footnotesize}
\end{program}


The division program can be refined to reduce the time of execution
by substituting repeated subtractions with a single subtraction of the product of the divisor and an estimated partial quotient 
to attain a partial quotient.
The new method is presented in program ``division02'' (program 9 and 10) and illustrated in Fig. 2(b). 
Since the product may be one digit longer than the divisor, to simply the programming,
each first element of the arrays is assigned with zero, and the partial dividend is always partitioned into the first $2:Ny+1$ elements.

The algorithms are as follows.

Step 1:
The dividend and divisor are
compared with each digit of the other number to determine which is larger. The leading few digits of the dividend or divisor are used to estimate the partial quotient.

Step 2:
Multiply the estimated partial quotient by the divisor and compare the value thus obtained with the partial dividend to determine whether the estimated partial quotient is too large.
If the estimated quotient is too large, then `1' is subtracted from the estimated quotient repeatedly until its value becomes equal to the dividend. This value is near the exact quotient.

Step 3:
Multiply the divisor by the newly estimated partial quotient using one of the aforementioned decomposing methods for multiplication and subtract the result from the partial dividend.
The difference is used to recalculate the partial dividend. Then return to Step 1 to compute the next digit of the quotient.

     \begin{program}
 \begin{footnotesize}
 \begin{verbatim}

PROGRAM division02
  implicit none
  integer :: ii, jj, k, kk, C, T, L=10, m=1
  integer, parameter :: ss=200, s=30 
  integer :: Xori(ss)=0, X(ss)=0, Y(ss)=0, P(ss)=0, Q(s)=0
  integer, parameter :: Nx=7, Ny=4, Ex=2, Ey=-4, Etot=Ex-Ey

  Xori(1:Nx) = (/0, 3, 8, 8, 7, 5, 6/) 
  Y   (1:Ny) = (/0, 1, 2, 9/)             

  do ii = 1, Ny
     X(ii) = Xori(ii)
  enddo 

  k = 1   ! k=1/2 denotes single integer/1st decimal place 

100 ii = 1


! Compare each digit of divident X and divisor Y and 
! calculate partial quotient when X(ii)>Y(ii).
! E.g.: X/Y=45316/742, X(ii)=453 < Y(ii)=742; 
! set Xori(4)=1 such that X(ii)=4531 > Y(ii)=742.

Loop1: DO WHILE (.true.)  
          if (X(ii)>Y(ii)) then 
             Q(k) = ( X(1)*10+X(2) ) / Y(2)   ! Q(1)=45/7
             exit
          else if (X(ii)==Y(ii)) then 
             ii = ii + 1
          else if (X(ii)<Y(ii)) then 
             do kk = 1, Ny
                X(kk) = X(kk+1)   
             enddo 
             X(Ny) = Xori(Ny+m)   ! 4530 -> 4531
             Q(k)  = 0             
             k = k + 1              
             m = m + 1 
             ii = 1    ! Reset ii to re-compare X and Y.
          endif                 
       END DO Loop1      

! Multiply estimated partial quotient by divisor. 
! 742*6 -> [4,4,5,2], P(Ny)=2, P(Ny-1)=5.

       C = 0
Loop2: do ii = Ny, 1, -1
          T = Y(ii)*Q(k) + C   
          C = T / L
          P(ii) = mod(T, L)    
       enddo Loop2

! Re-calculate partial quotient.

       ii=1
Loop3: DO WHILE(.true.)
         if (P(ii)>X(ii)) then     ! E.g.,5194>4531,Q=Q-1
            Q(k) = Q(k)-1 
     Loop4: do ii = Ny, 1, -1      ! Recalculate P.
               T = Y(ii)*Q(k) + C
               C = T / 10
               P(ii) = mod(T,L)     
            enddo Loop4
         else if (P(ii)==X(ii)) then
            ii = ii+1
         else if (P(ii)<X(ii)) then ! E.g.,4452<4531,exit.
            exit
         endif
       END DO Loop3
\end{verbatim}
\caption{Division of Two Floats - Method 2} 
\end{footnotesize}
\end{program}

     \begin{program}
 \begin{footnotesize}
 \begin{verbatim}


! Subtract P from dividend X and return remainder to X.

Loop5: do ii = Ny, 1, -1            
          if (P(ii)>X(ii)) then        ! X-P=4531-4452=0079
             X(ii) = X(ii) + 10 - P(ii)
             X(ii-1) = X(ii-1)- 1
          else if (P(ii)<=X(ii)) then  ! Subtract directly.
            X(ii) = X(ii)-P(ii)
          endif
       enddo Loop5
       if (X(1)==0) then     
Loop6:    do kk = 1, Ny
             X(kk) = X(kk+1) 
          enddo Loop6
          X(Ny) = Xori(m+Ny)           ! Reset Xori.
       endif
       k = k + 1   ! Calculate next partial quotient.
       m = m + 1
       IF ( k <= s ) GOTO 100
1000   print*, " quoitent:"
       write(*, "(1I1, 1H., 29I1, 7H * 10^ , 1I4)") &
             Q(1), (Q(ii), ii =2, s), Etot
END PROGRAM
\end{verbatim}
\caption{Division of Two Floats - Method 2 Continued} 
 \end{footnotesize}
\end{program}

   \begin{figure}
\centering
(a) 
    {
         \includegraphics[width=2.4in]{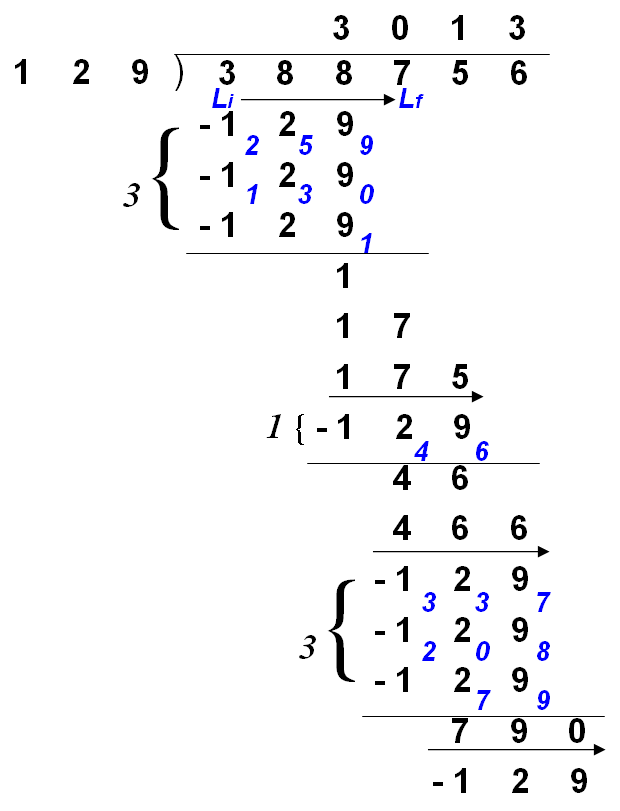}

    }
   \vskip 0.3in
     {
      \includegraphics[width=3.4in]{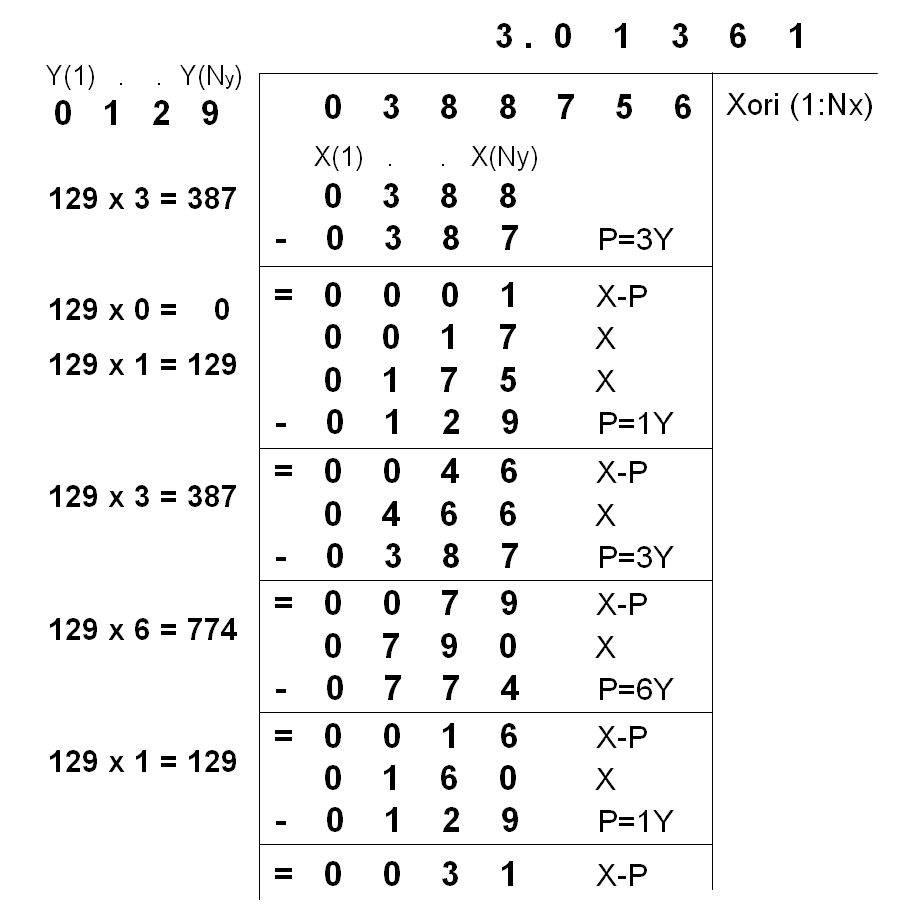}  
   \vskip 0.3in
(b)
     }
\\
   \vskip 0.3in
   \caption{Two Repeated Subtraction Methods for Division of Floats.}
\end{figure}


\section{Validity of Stirling Approximation}


The factorial function is related to Stirling's formula [2], 
\begin{equation}
 \lim_{n \rightarrow \infty} \frac{n!} {\sqrt{2\pi n} \ n^{n} e^{-n}}= 1,
\end{equation}
and its logarithmic from -- Stirling approximation:
\begin{equation}
\begin{array}{ll}
 ln(n!) & =   \sum_{k=1}^{n} ln(k)
      \approx \int_{x=1}^{x=n} ln(x) dx   \\
      &  =  [x\ ln(x)-x]_1^{n} = n \ ln(n)- n + 1 \\
      &  \approx   n \ ln(n)- n 
\end{array}
\end{equation}
Eq (2) is a simple and direct proof of Stirling's approximation without invoking Stirling's formula,
whose associated error can be obtained
using the gamma function $\Gamma(n)=n!=\int_0^{\inf} e^{-x} x^{n} dx$,
for both integer and non-integer.
Since $n!$ are very large even for relatively small values of $n$,
Stirling approximation is more often used.
References states that the equation for sufficiently large $n$ from Stirling formula,
$
n! \approx \sqrt{2 \pi n} \ n^{n} e^{-n},
$
works well for small $n$;
at $n=10$ the error is $\sim 1\%$.
However, the acceptable limit of $n$ to Stirling approximation
is not clear.

We have computed the number of configurations $W$ for 
$n$=$100$ distinguishable objects being distributed into 4 identical states,
with each state containing \{$n_{1}$,$n_{2}$,$n_{3}$,$n_{4}$\}=\{10,20,30,40\} objects.
If Stirling approximation is applied to the solution
$W$=$n!/n_{1}!n_{2}!n_{3}!n_{4}!$,
such that
$ln W$ = $n ln(n)$ - $\Sigma_{i} n_{i} ln(n_{i})$,    
then the error will exceed $100\%$.
While the large error is due to misuse of the approximation equation,
the precise value of $W$ can be computed using the aforementioned methods.
We found that $n \ge 100$ is needed for $ln(n!)/(n\ln(n)-n)$ to attain an error of $\sim 1\%$.


\section{Application of High Precision Technique to the Engineering Industry and Scientific Computing}

Technique to achieve high precision arithmetic 
is commonly used in 
public-key cryptography,
which requires the manipulation of integers of thousands of digits,
and also experimental mathematics that involves the computation of fundamental mathematical constants, such as pi.
The high precision scheme is also essential to
the digital design of firmware in 
digital signal processing systems,
including cell phone, PDAs, and GPS,
which use a specialized microprocessor, the digital signal processor (DSP) [3], 
to process input signals with very limited memory or cache.
 When DSPs receive too many input signals exceeding
 the limit of cache or memory,
 the two's complement method was adopted 
to store a number of up to $2n$ bits using $n$ bits.
Most DSPs are based on fixed point arithmetic
 because the dynamical range provided by the floating point is not needed for signal processing.
In the future DSPs may be adopted to reduce the cost and complexity of software development,
at the expense of more expensive hardware,
to attain a wider dynamic range like the floating point DSPs which are often used in military signal applications
involving Radar, Sonar, and communications..


There is a rapidly growing need
for a higher level of numeric precision in scientific computing.
Recently, Bailey and Borwein 
(2008) conducted a survey of the high-precision package applied to scientific computing [4]. 
One of the application was
supernovae with the QD package, 
which provides double-double (128-bit or 31-digit) and quad-double (256-bit or 62-
digit) datatypes to solve for the non-local thermodynamic equilibrium populations
of iron and other chemical elements [5]. 
 Iron may be present in the outer envelope
of a
supernova as Fe II (Fe+1), but in the inner parts Fe IV (Fe+3) or Fe V (Fe+4) could be dominant.
Since the relative population of any state
is proportional to the exponential of the ionization energy, the
dynamic range of these numerical values can be very large.
Introducing artificial cutoff for the elemental distribution leads to spurious numerical glitches.
In order to solve for all of the populations and yet reduce the computing time,
 a scheme was employed to determine whether to use 64-bit, 128-bit or 256-
bit arithmetic in constructing and solving the linear system.

In addition to supernova simulations,
Bailey \& Borwein's survey
shows that the applications
comprise climate modeling, planetary orbit calculations, Coulomb n-body atomic systems, 
scattering amplitudes of elementary particles,
nonlinear oscillator theory, Ising theory, quantum field theory, as well as experimental mathematics.
High-precision arithmetic techniques are becoming indispensable
to modern large-scale computational science.

\section {Discussion and Conclusions}

The approach of calculating exact numbers that exceed a  machine's limit has been utilized in the electronic industry, 
and is becoming important to scientific computing. 
However, such a scheme is much less well recognized by researchers and instructors of scientific disciplines.
 Despite the fact that
  various arithmetical packages and stand-alone software
 are available for performing high-precision calculations, 
 one of the most popular is the GMP (Gnu Multiple Precision Arithmetic) library, which is written in C,
and Mathematica,
which along with several machine-code languages adopts GMP,
documents for the underlying algorithms have been scarce until only recently 
[6]. 
Motivated by the need to improve the educational value of computational physics courses, we performed an independent study on 
basic arithmetic operations to arbitrary precision.
Various computing algorithms for the multiplication and division of overflowing/underflowing numbers are explored, which can be integrated to solve complicated algebraic equations.

Notably,
software packages  
to solve
mathematical functions as fundamental as the natural logarithm are still either under development or unavailable (as in GMP).
Complicated computations using the multiple precision 
routines may suffer precision loss during intermediate calculations,
depending on the problems and the size of computer memory, 
making the last few digits of the result unreliable.
%
Understanding the approach and thus the imposed limitation is important
to assure accurate solutions.

The algorithms for multiplication and division to an arbitrary precision are summarized as follows.

1. Given two numbers $X$ and $Y$, to calculate the product $P=X*Y$ or the quotient $Q=X/Y$, where $X$ and $Y$ could exceed the machine number limit, the digits of X and Y are arranged into an array of elements: $X=[X(Nx),.,X(2),X(1)]$ and $Y=[Y(Ny),.,Y(2),Y(1)]$;
and $P=[P(Nx+Ny),.,P(2),P(1)]$. The multiplication of an $Nx$-digit with an $Ny$-digit number can generate a number with either $Nx+Ny$ or $Nx+Ny-1$ digits. The precision achieved for division is determined by the declared array size of the quotient.

2. If a 32-bit fix-point number is used, then one array element can accommodate ten digits. To prevent overflow during the computation (such as multiplication), each element can hold a number of four digits. Allocating more digits to an element consumes less memory and improves performance.
However, the costs in terms of memory and CPU time of our problem are not excessive, so one digit can be allocated to each element.

3. The algorithm for multiplication is based on decomposing the multiplicand, while the algorithm for division is based on decomposing the dividend. For multiplication, extra digits of intermediate products are carried to higher places and accumulated to yield the final product. The algorithm for division is much more difficult: a set of numbers is extracted to form a partial dividend, from which the divisor is repeatedly subtracted. These subtractions yield a partial quotient. This procedure continues to lower places and the total quotient is identical to that obtained by dividing the dividend by the divisor once.

4. Given the exponents to the fixed numbers, the multiplication and division of these fixed numbers can be extended straightforwardly to floating-point arithmetic.

5. The method can be regarded as an extension of the binary system in which 1 bit has one of two possible values and the decimal system in which 1 digit has one of ten values to a large-number system such as one in which an element may comprise 10000 to 1000000000 values.

Given the subtraction method of division,
numerous mathematical functions as simple as $x^{3.0}=100$ and algebraic equations of the form $f(x)=0$
can be solved to a high precision via Newton-Raphson iteration, $x_{i+1}=x_{i}-f(x_{i})/f'(x_{i})$,
provided the derivative of the function is known. However, for mathematical functions such as logarithms, series approximations must be employed to find the roots [7]. 
Several computer algebra systems that can express almost any number, including Mathematica and Maple, combine the arbitrary-precision scheme with symbolic computation, which transforms mathematical expressions into symbolic form before calculations are made, unlike truly numerical and, therefore, limited-precision computation.

For memory-extensive computations,
memory can be allocated and released via the allocatable array,
which can be controlled more effectively by C (using the malloc and memset functions) than by Fortran. 
Such a case may involve a large linearized system. For example, the implicit scheme
may be invoked to solve a 
set of Navier-Stokes equations in multi dimensions.
Methods for constructing and solving such a complex system of equations with a high precision are still being developed [4].

Studies of the newly emerged General-Purpose computing on a Graphical Processing
Unit (GUGPU), which uses a multi-thread GPU as a co-processor with the CPU,
demonstrate that PDE problems are solved much faster using GPUs than using CPUs.
GPUs had been restricted to single precision floating point arithmetic until
 double precision GPU architecture (NVidia Fermi GPU) became available in
2010. Before then, mixed precision iterative solvers for finite element simulations
had been implemented by iteratively computing the residuals of a single precision approximate solution to acquire a higher precision, or defect correction, assigning the computationally intensive part to the less precise GPU and the precise correction part to the CPU [8]. However, for small- to medium-size problems, the GPU-CPU approach to increasing precision is not always faster than the pure CPU approach, because the CPU with a large cache outperforms the GPU. Emulating double-precision arithmetic on the single-precision GPU typically increases tenfold the operation count [9]. GPU programming requires complicated memory management which involves shared and global memories between threads and registers to implement effectively parallel algorithms on various hardware levels. The implementation of arithmetic to arbitrary precision on a GPU has not been
documented, to the best of our knowledge.


The arbitrary precision algorithms for multiplication and division that are presented
herein can be integrated to solve advanced mathematical problems to
multiple precision or almost arbitrarily-high precision, limited only by the
memory of the host machine.
Notably, the success of the arbitrary precision implementation depends on the smart
use of arrays and careful programming,
and the details of the implements can vary greatly from problem to problem,
or even within one problem.
Finally,
as ill-conditioned matrices
are inevitably present in a system of implicitly discretized PDEs,
where a small error in the matrix leads to a large error in the solution,
the arbitrary precision technique, even for the GUGPU, may become an important future exploration in computing.

*{\bf Acknowledgments}
We are grateful to the three referees 
for helpful comments on the manuscript.
Special thanks are due for the anonymous referee \#2 
for his valuable comments on an early version of this work submitted to Computer Physics Communications for the special issue of CCP2009.
We thank Drs. David Bailey, Fang-An Kuo and Shuen-Tai Wang for useful correspondence on the multiple precision and GPU computing.
 Yan-Hou Chang is appreciated for his help in drawing figure. 
 Ted Knoy is appreciated for his editorial assistance.
This work is supported by the National Science Council and the National Center for High-Performance Computing of Taiwan.


\begin{thebibliography}{10}
 \bibitem{paperprecision}
Goldberg, D. (1991) `What Every Computer Scientist Should Know About Floating-Point Arithmetic', 
{\it ACM Computing Surveys}, Vol.~23, pp.154--225. 

 \bibitem{paperStirling}
Brandt, K. A. \&  Wallner A. S. (1999)
`The Validity of Stirling's Approximation: A Physical Chemistry Project',
{\it J. Chem. Educ.}, Vol.~76, pp.1395--1397



\bibitem{paperDSP}
Grover D. `Where Do I Learn DSP? Dale Grover', 
http://www.redcedar.com/learndsp.htm





\bibitem{paperSurvey}
Bailey, D. H. \& Borwein, J. M. (2008) 
`High-Precision Computation and Mathematical Physics', 
{\it XII Advanced Computing and Analysis Techniques in Physics Research},
available at http://crd.lbl.gov/$\sim$dhbailey/dhbpapers/dhb-jmb-acat08.pdf.


\bibitem{paperSupernova}
Hauschildt, P. H. \& Baron, E. (1999) 
`The numerical solution of the expanding Stellar atmosphere problem',
{\it J. Comp. and Applied Math.}, Vol.~109, pp.41--63.





\bibitem{paperComputerArith}
Brent, R. P. \& Zimmermann, P. (2010) `Modern Computer Arithmetic', book manuscript,
version 0.5.7, http://www.loria.fr/$\sim$zimmerma/mca/mca-cup-0.5.7.pdf. 


\bibitem{paperMath}
Bailey, D. H.,
`Experimental Mathematics and High-Performance Computing',
http://crd.lbl.gov
/$\sim$dhbailey/dhbtalks/dhb-expmath-hpc.pdf.









\bibitem{papergpu}
Goddeke, D., Strzodka, R. \& Turek, S. (2005)  
`Accelerating Double Precision FEM Simulations with GPUs',
{\it Proceeding of ASIM 2005 -- 18th Symposium on Simulation Technique}.

\bibitem{paperGaia}
Meredith, J. `Emulated double precision in the Gaia project at the Lawrence Livermore National Laboratory',
http://www.ll.mit.edu/HPEC/agenda
/proc04/powerpoints/Talks-OPEN/Tues
/Johnson.ppt.

\end{thebibliography}
\end{document}